\documentclass[lettersize,journal, 10pt]{IEEEtran}
\usepackage{amsmath,amsfonts}
\usepackage{algorithm}
\usepackage{array}
\usepackage[caption=false,font=normalsize,labelfont=sf,textfont=sf]{subfig}
\usepackage{textcomp}
\usepackage{stfloats}
\usepackage{url}
\usepackage{verbatim}
\usepackage{graphicx}
\usepackage{cite}

\usepackage{tikz}
\usepackage{physics}
\usepackage{eucal}
\usepackage{xspace}
\usepackage{multirow}
\usepackage{algorithm}
\usepackage{algpseudocode}
\usepackage{enumitem}

\usepackage{soul}
\usepackage{color}
\usepackage{xcolor}
\usepackage{booktabs} 
\usepackage{fancyhdr}

\usepackage{amsmath,amssymb,amsfonts}
\usepackage{graphicx}
\usepackage{textcomp}
\usepackage{cite}

\usepackage{hyperref}
\usepackage{cleveref}

\hypersetup{
    colorlinks = true,
    linkcolor = blue,
    urlcolor  = blue,
    citecolor = blue,
}

\newcommand{\Zbasis}{\textrm{Z}-basis\xspace}
\newcommand{\Xbasis}{\textrm{X}-basis\xspace}
\newcommand{\Ybasis}{\textrm{Y}-basis\xspace}
\newcommand{\mYbasis}{\textrm{-Y}-basis\xspace}

\BeforeBeginEnvironment{equation}{\begingroup\small\vspace{-5pt}}
\AfterEndEnvironment{equation}
{\endgroup}

\BeforeBeginEnvironment{equation*}{\begingroup\small\vspace{-5pt}}
\AfterEndEnvironment{equation*}
{\endgroup}

\BeforeBeginEnvironment{align}{\begingroup\footnotesize\vspace{-5pt}}
\AfterEndEnvironment{align}
{\endgroup}

\BeforeBeginEnvironment{align*}{\begingroup\footnotesize\vspace{-5pt}}
\AfterEndEnvironment{align*}
{\endgroup}

\usepackage{etoolbox}
\AtBeginEnvironment{table}{\vskip-0.3em}

\begin{document}

\title{Enhance Quantum Teleportation with Multi-Axis Measurement}

\author{
Junyao Zhang\textsuperscript{1}, 
Jonathan Ku\textsuperscript{1},
Zhiding	Liang\textsuperscript{2},
Hai Li\textsuperscript{1},
Yiran Chen\textsuperscript{1},
Ben	McCarty\textsuperscript{3}\\
\normalsize{
\textsuperscript{1}Duke University, 
\textsuperscript{2}Rensselaer Polytechnic Institute,
\textsuperscript{3}Accenture}
}




\maketitle

\begin{abstract}

Quantum teleportation is a cornerstone of quantum information processing, enabling the nonlocal transmission of quantum states across arbitrary distances using shared entanglement and classical communication. While the standard protocol typically employs \Zbasis Bell-state measurements, this fixed-basis approach limits flexibility in practical quantum networks, where dynamic operations, hardware variability, and advanced communications demand alternative measurement bases.

In this work, we introduce a multi-axis quantum teleportation protocol that generalizes the measurement process by allowing arbitrary basis choices. We provide a formal derivation and self-contained mathematical proof demonstrating that the input quantum state can be faithfully reconstructed under basis-adaptive restoration operations. By establishing a rigorous algorithmic and analytical foundation, this work validates the generalized teleportation protocol and paves the way toward advanced strategies for quantum communication. 
The demonstrations of the proposed protocol are available at: 
\href{https://github.com/JJJayyyy/Multi-Axis-Quantum-Teleportation}{multi-axis QT}.

\end{abstract}


\section{Introduction}

Quantum teleportation~\cite{qt} is one of the foundational protocols in quantum information processing~\cite{quantum_internet}. It enables one party (Alice) to transfer an unknown quantum state to another party (Bob) across arbitrary distances by leveraging shared entanglement (Einstein-Podolsky-Rosen (EPR) pair) and classical communication. The protocol relies on the interplay of quantum entanglement and the no-cloning theorem~\cite{non_clone}, ensuring that quantum states can be transferred without violating fundamental principles of quantum mechanics and bypassing the risk of losing the qubit to transmission loss \cite{wehner_quantum_2018}. 
Over the past decades, quantum teleportation has been experimentally realized in multiple physical platforms, ranging from photonic qubits to atom-photon hybrids~\cite{qt_photon, hybrid_q_photon, qt_progress}, and over distances ranging from laboratory-scale setups to satellite links~\cite{qt_1400, qt_dqc, hermans_qubit_2022}.
Satellite-based experiments have shown their potential for global quantum connectivity, enabling reliable quantum state transfer over long distances despite significant channel losses. At shorter scales, teleportation supports distributed quantum computing by allowing remote processors to exchange qubits as if locally connected~\cite{qt_dqc}. These advances demonstrate the versatility of teleportation for quantum networks, distributed quantum computation, and secure communication~\cite{q_info, quantum_internet, qt_dqc}.

Although quantum teleportation is often described abstractly, its physical implementation requires the parties involved to Bell-state measurement \cite{exp_qt, bell_measurement}.
A commonly used measurement is performed in the Z-basis by Alice on her half of the EPR pair and on the message qubit to be teleported \cite{
mor1999teleportation, gordon2006generalized}. This Bell-state measurement projects her two qubits onto one of four maximally entangled states, causing Bob’s qubit (the other half of the EPR pair) to collapse into a state correlated with Alice’s measurement result. Depending on the Bell-state measurement outcome, Bob applies corresponding restoration operations to recover the original input state. 
However, in practical quantum network scenarios, Z-basis measurements alone are insufficient for ensuring the flexibility and robustness required by advanced communication tasks. Applications such as quantum repeaters, fault tolerance design, multi-party communication, and entanglement swapping often demand measurements in various Pauli or Bell bases~\cite{QFI_qt_2, quantum_internet, qce_qt}. More generally, Positive Operator-Valued Measures (POVMs) extend beyond orthogonal bases and enable conclusive teleportation, where perfect state transfer is possible with non-unit success probability~\cite{mor1999teleportation}. Parameterizing such bases can optimize fidelity and success rates~\cite{gordon2006generalized}. Importantly, any change in measurement basis (or POVM) necessitates a corresponding adaptation of Bob’s restoration circuit, situating our multi-axis scheme within this broader class of generalized teleportation protocols.

To address this challenge, we propose a multi-axis measurement scheme designed to streamline and generalize these measurements and restoration operations within the teleportation protocol. By allowing qubits to be measured along different axes with minimal overhead, the scheme addresses common issues such as randomness in entanglement distribution, hardware inefficiencies, and the need for rapid basis changes \cite{multi_measurement, multi_measurement_1, zhao2004experimental, multi_measurement_2}. 
We demonstrate how our approach can be integrated into existing teleportation frameworks, thereby improving quantum networks' scalability and reliability.

In the following sections, we formally derive the state transfer equivalence under various measurement bases and provide a complete mathematical proof demonstrating that the input state can be faithfully reconstructed through appropriately modified restoration operations. We also discuss the implications for secure communication, where basis diversity can enhance the resilience of quantum cryptographic protocols.

\section{Preliminary}

\subsection{Quantum Teleportation} \label{sec:qt}
Quantum teleportation is a foundational protocol in quantum information science, enabling the transfer of an arbitrary quantum state from one location to another using both quantum and classical resources.
\begin{figure}[htbp]
    \centering
    \includegraphics[width=\linewidth]{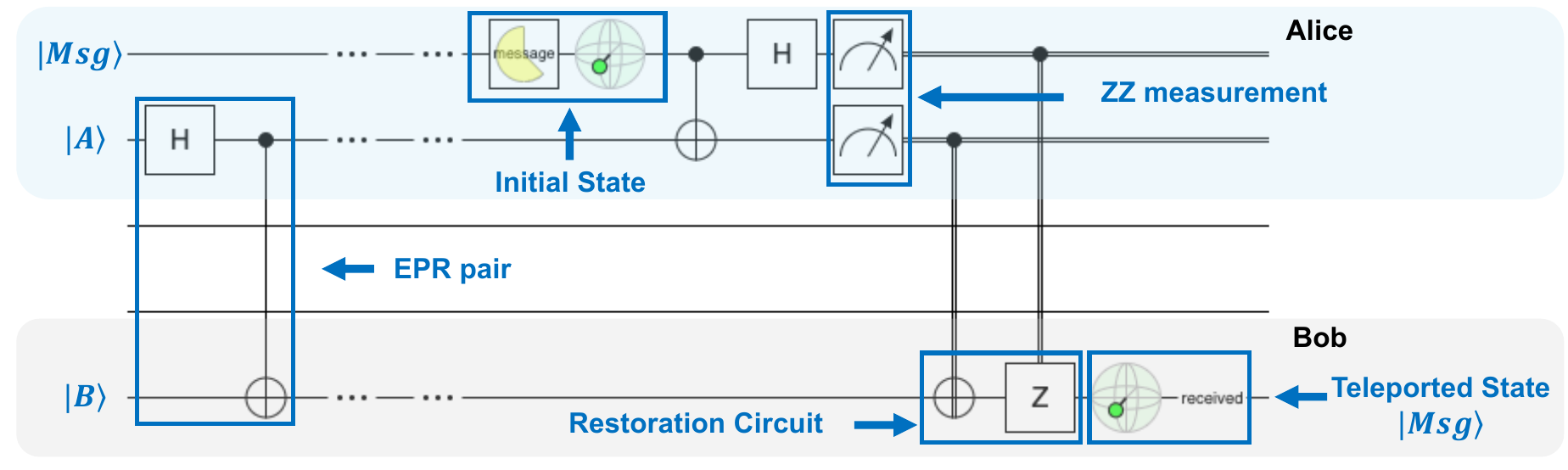}
    \caption{Quantum teleportation protocol where both $\ket{\text{Msg}}$ and $\ket{\text{A}}$ are measured in the \Zbasis. Bob recovers the original message state with restoration: $\{\mathrm{I}, \mathrm{X}, \mathrm{Z}, \mathrm{ZX}\}$.}
    \label{fig:QT}
\end{figure}
\cref{fig:QT} illustrates the quantum circuit for this protocol. For conceptual clarity, the protocol is represented using standard quantum-circuit notation; In practical implementations, particularly in photonic systems, these operations can be physically realized using optical components such as wave plates. In this setup, Alice and Bob each hold one half of an entangled Einstein–Podolsky–Rosen (EPR) pair, which serves as the quantum communication channel. The EPR pair is initialized in the Bell state:

\begin{align*}
    \ket{\text{Bell}} = \frac{\sqrt{2}}{2}(\ket{0}_A\ket{0}_B + \ket{1}_A\ket{1}_B)
\end{align*}

\noindent
Here, $\ket{\text{A}}$ and $\ket{\text{B}}$ denote Alice's and Bob’s respective halves of the entangled pair. Let $\ket{\text{Msg}} = \alpha\ket{0} + \beta\ket{1}$ be the single-qubit message state that Alice intends to send to Bob:

\begin{align*}
    \ket{\text{Msg}} = 
\begin{bmatrix} 
    \alpha \\
    \beta  \\
\end{bmatrix}
= \alpha\ket{0} + \beta \ket{1}
\end{align*}

\noindent
The complete system therefore consists of three qubits: two under Alice’s control, $\ket{\text{Msg}}$ and $\ket{\text{A}}$ (shown in blue in \cref{fig:QT}); One under Bob’s control, $\ket{\text{B}}$ (shown in gray in \cref{fig:QT}). The combined state of the three-qubit system is then given by:

\begin{align*}
\ket{\text{Msg}}\otimes\ket{\text{Bell}} 
 &   = \frac{\sqrt{2}}{2}\Bigl[\alpha\ket{0}\otimes(\ket{00}+\ket{11})+\beta\ket{1}\otimes(\ket{00}+\ket{11})\Bigr]\\
 &   = \frac{\sqrt{2}}{2}(\alpha\ket{000}+\alpha\ket{011}+\beta\ket{100}+\beta\ket{111})
\end{align*}

\noindent 
To teleport the message, Alice entangles message qubit $\ket{\text{Msg}}$ with her half of the EPR pair $\ket{\text{A}}$ by applying a CNOT gate, followed by a Hadamard gate, in accordance with the protocol. The system state evolves into the following superposition:

\begin{align*}
&  (\text{H}\otimes \text{I} \otimes \text{I}) (\text{CNOT}\otimes \text{I})(\ket{\text{Msg}}\otimes\ket{\text{Bell}})\\
 &   =(\text{H}\otimes \text{I} \otimes \text{I})(\text{CNOT}\otimes \text{I})\frac{\sqrt{2}}{2}(\alpha\ket{000}+\alpha\ket{011}+\beta\ket{100}+\beta\ket{111})\\
 &   =(\text{H}\otimes \text{I} \otimes \text{I})\frac{\sqrt{2}}{2}(\alpha\ket{000}+\alpha\ket{011}+\beta\ket{110}+\beta\ket{101})\\
 &   = \frac{1}{2}\Bigl[\alpha(\ket{000}+\ket{011}+\ket{100}+\ket{111}) \\
 & \qquad +\beta(\ket{010}+\ket{001}-\ket{110}-\ket{101})\Bigr]
\end{align*}

\noindent
Rewriting expression by grouping Alice’s two qubits yields:

\begin{align*}
& (\text{H}\otimes \text{I} \otimes \text{I}) (\text{CNOT}\otimes \text{I})(\ket{\text{Msg}}\otimes\ket{\text{Bell}})\\
& = \frac{1}{2}\Bigl[
     \ \ket{00}(\quad\alpha\ket{0}+\beta\ket{1})\\
     & \qquad + \ket{01}(\quad\beta\ket{0} + \alpha\ket{1})\\
     & \qquad + \ket{10}(\quad\alpha\ket{0}-\beta\ket{1})\\
     & \qquad + \ket{11}(-\beta\ket{0}+\alpha\ket{1})\Bigr]
\end{align*}

\noindent 
Alice then performs a joint measurement (denoted as $\mathrm{ZZ}$ measurement in \cref{fig:QT}) on her two qubits ($\ket{\text{Msg}}$ and $\ket{\text{A}}$). 
This measurement collapses her qubits into one of four orthogonal basis states: $\ket{00}$, $\ket{01}$, $\ket{10}$, or $\ket{11}$, and yields two classical bits. Importantly, this process irreversibly destroys the original quantum state in Alice’s possession, adhering to the no-cloning theorem~\cite{non_clone}, and collapses Bob’s qubit into one of the following conditional states:

\begin{align*}
     & \ket{00}\rightarrow(\quad\alpha\ket{0}+\beta\ket{1})\\
     & \ket{01}\rightarrow(\quad\beta\ket{0} + \alpha\ket{1})\\
     & \ket{10}\rightarrow(\quad\alpha\ket{0}-\beta\ket{1})\\
     & \ket{11}\rightarrow(-\beta\ket{0}+\alpha\ket{1}))
\end{align*}

\noindent
Alice then sends the two classical bits to Bob over a classical channel. Upon receiving them, Bob determines which of the four possible states his qubit has collapsed into. He then applies the corresponding unitary operation (denoted as restoration circuit in \cref{fig:QT}) to recover the original message state. The mapping from classical bits to restoration operations is summarized below:

\begin{table}[h]
  \centering
  \begin{tabular}{ccc}
    Bits from Alice     & Bob's State                       & Operation \\
    \midrule
    $\mathbf{00}$       & $\alpha\ket{0}+\beta\ket{1}$      & $\mathbf{I}$\\
    $\mathbf{01}$       & $\beta\ket{0} + \alpha\ket{1}$    & $\mathbf{X}$\\
    $\mathbf{10}$       & $\alpha\ket{0}-\beta\ket{1}$      & $\mathbf{Z}$\\
    $\mathbf{11}$       & $-\beta\ket{0}+\alpha\ket{1}$     & $\mathbf{ZX}$
  \end{tabular}
\end{table}

\noindent
After applying the appropriate operation, Bob’s qubit $\ket{\text{B}}$ is restored to the original state $\alpha\ket{0}+\beta\ket{1}$. The entanglement resource is consumed in the process, and the teleported state now resides solely with Bob, completing the quantum teleportation protocol without duplicating the quantum information.

\subsection{Measurements in Different Axis}\label{sec:multi_axis}
A foundational concept in quantum computing is measurement, where one reads information from a qubit state. Typically, this measurement is performed in the \Zbasis, which is formed by the orthonormal states

\begin{align*}
    \ket{0} = \begin{bmatrix} 1 \\ 0 \end{bmatrix}, 
    \quad
    \ket{1} = \begin{bmatrix} 0 \\ 1 \end{bmatrix}.
\end{align*}

In quantum computing frameworks \cite{multi_axis_0, Qiskit}, measurements are implemented primarily in the computational \Zbasis. However, measurement need not be restricted to this basis. For example, to measure in the $\pm \mathrm{X}$ basis, one can first apply a Hadamard (H) gate that maps $\ket{0} \mapsto \ket{+}$ and $\ket{1} \mapsto \ket{-}$. A subsequent \Zbasis measurement then effectively measures in the \Xbasis. More generally, any pair of orthogonal states $\{\ket{\psi_0}, \ket{\psi_1}\}$ constitutes a valid measurement basis in Hilbert space. To realize measurement in this arbitrary basis, one applies an appropriate unitary transformation that maps $\ket{\psi_0}$ to $\ket{0}$ and $\ket{\psi_1}$ to $\ket{1}$, followed by the \Zbasis measurement. This more flexible approach is referred to as \textbf{Multi-Axis measurement} or measurement in different bases~\cite{q_info, multi_axis_0}.

The choice of measurement basis is ultimately dictated by the observable of interest and the experimental setup. Different measurement bases probe different aspects of the qubit state, underscoring the fundamental distinction between quantum bits and their classical counterparts \cite{non_clone}. Such flexibility in measurement is central to many quantum algorithms and protocols, allowing for the extraction of diverse information from the same underlying quantum system \cite{Kundu2024MeasurementDependence, zhao2004experimental, multi_measurement_2}. We summarize several commonly used bases and their corresponding unitary transformation, including $\{\mathrm{X}, \mathrm{Y}, -\mathrm{Y}\}$ bases~\cite{multi_axis_0}

\begin{table}[H]
  \centering
  \begin{tabular}{ccc}
    Base     &  Transformation   \\
    \midrule
    $\mathrm{X}$   & $\mathrm{H}$ \\
    $\mathrm{Y}$   & $\mathrm{S}^\dagger \mathrm{H}$ \\
    $\mathrm{-Y}$  & $\mathrm{SH}$   
  \end{tabular}
  \label{tab:base}
\end{table}

\noindent
The corresponding unitary operators are defined as follows:

\begin{equation*}
    \mathrm{H} = \tfrac{1}{\sqrt{2}}
    \begin{pmatrix}
        1 & 1 \\
        1 & -1
    \end{pmatrix},
    \quad
    \mathrm{S} = 
    \begin{pmatrix}
        1 & 0 \\
        0 & i
    \end{pmatrix},
    \quad
    \mathrm{S}^\dagger =
    \begin{pmatrix}
        1 & 0 \\
        0 & -i
    \end{pmatrix}
\end{equation*}

\section{Quantum Teleportation with Multi-Axis Measurements}

Building on the standard teleportation procedure mentioned earlier, we extend the procedure by introducing measurements in alternative bases.
These enhancements improve the flexibility and robustness of quantum teleportation, enabling it to accommodate more general quantum network settings. In this section, we use the \Ybasis as an example to demonstrate that the modified teleportation protocol can faithfully transmit an arbitrary quantum state to the intended receiver.

\subsection{\texorpdfstring{Teleportation with Y Measurement on Both $\ket{\text{Msg}}$ and $\ket{\text{A}}$}{Teleportation with Y Measurement on Msg and A}} \label{sec:YY}

As introduced in \cref{sec:multi_axis}, measuring in the \Ybasis can be achieved by applying an $\mathrm{S}^\dagger$ gate followed by a Hadamard ($\mathrm{H}$) gate before the standard \Zbasis measurement. \cref{fig:QT_YY_error} shows the quantum circuit of this YY-measurement setup, simulated using \cite{quirk}. The state of the system is thus transformed to:

\begin{figure}[b]
    \centering
    \includegraphics[width=\linewidth]{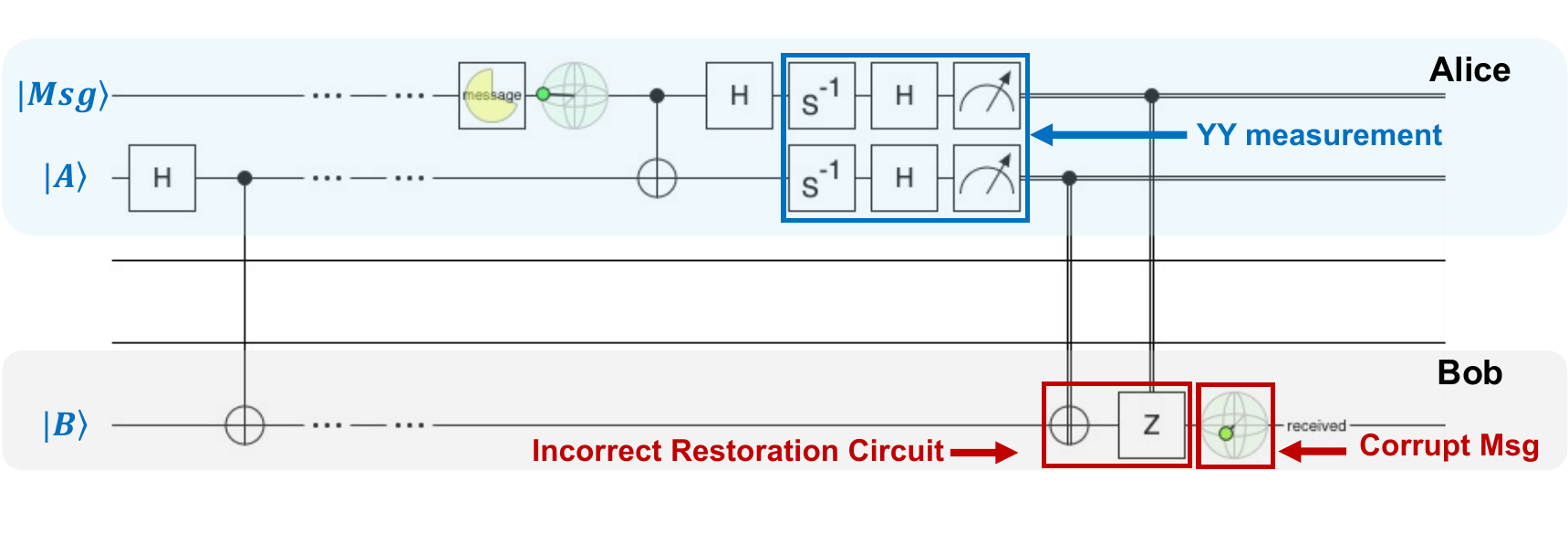}
    \caption{Quantum teleportation protocol where both $\ket{\text{Msg}}$ and $\ket{\text{A}}$ are measured in the $\mathrm{Y}$ basis. Bob fails to recover the original message state with original restorations $\{\mathrm{I}, \mathrm{X}, \mathrm{Z}, \mathrm{ZX}\}$.}
    \label{fig:QT_YY_error}
\end{figure}

\begin{align*}
 &   (H \otimes H \otimes I)(S^\dag \otimes S^\dag \otimes I)(H\otimes I \otimes I)(\text{CNOT}\otimes I)(\ket{\text{Msg}}\otimes\ket{\text{Bell}})\\
 &   =(H \otimes H \otimes I)(S^\dag \otimes S^\dag \otimes I)(H\otimes I \otimes I)(\text{CNOT}\otimes I)\frac{\sqrt{2}}{2}(\alpha\ket{000} \\
 & \qquad + \alpha\ket{011}+\beta\ket{100}+\beta\ket{111})\\
 & = \frac{1-i}{4}
 \Bigl[
    (\alpha+\beta)\ket{000} + i(-\alpha+\beta)\ket{001} + (\alpha-\beta)\ket{010} \\
 & \qquad\quad 
    +i(\alpha+\beta)\ket{011} + i(\alpha-\beta)\ket{100} + (\alpha+\beta)\ket{101} \\
 & \qquad\quad 
    +i(\alpha+\beta)\ket{110} + (-\alpha+\beta)\ket{111}\Bigr]
\end{align*}

\noindent Grouping the state by Alice’s measurement outcomes gives:

\begin{align*}
     = \frac{1-i}{4}\Bigl[
     & \quad\ket{00}((\alpha+\beta)\ket{0} + i(-\alpha+\beta)\ket{1})\\
     & + \ket{01}((\alpha-\beta)\ket{0} + i(\alpha+\beta)\ket{1})\\
     & + \ket{10}(i(\alpha-\beta)\ket{0} + (\alpha+\beta)\ket{1})\\
     & + \ket{11}(i(\alpha+\beta)\ket{0} + (-\alpha+\beta)\ket{1})\Bigr]
\end{align*}

\noindent 
Bob subsequently attempts to apply the standard restoration operators from the ZZ measurement case, $\{\mathrm{I}, \mathrm{X}, \mathrm{Z}, \mathrm{ZX}\}$, to recover $\ket{\text{Msg}}$. However, due to the change in measurement basis, these standard operations lead to incorrect results, as shown in \cref{fig:QT_YY_error}. For example, with classical bits $\mathbf{01}$:

\begin{align*}
     \mathrm{X} \cdot [(\alpha-\beta)\ket{0} + i(\alpha+\beta)\ket{1}]
    = \begin{pmatrix}
        0 & 1 \\
        1 & 0
    \end{pmatrix} 
    \begin{pmatrix}
        \alpha-\beta    \\
        i(\alpha+\beta) \\
    \end{pmatrix}
    = 
    \begin{pmatrix}
        i(\alpha+\beta) \\
        \alpha-\beta    \\
    \end{pmatrix}
\end{align*}

\noindent 
The results from the standard ZZ restoration for each classical bit outcome are:

\begin{table}[H]
  \centering
  \scriptsize
  \begin{tabular}{ccc}
     \textbf{Bits from Alice}   &  \textbf{Default Restoration} & \textbf{Bob's Post-Restoration State}\\
    \midrule
    $\mathbf{00}$  & $\mathbf{I}$  &  $(\alpha+\beta)\ket{0} + i(\alpha-\beta)\ket{1}$  \\
    $\mathbf{01}$  & $\mathbf{X}$  &  $i(\alpha+\beta)\ket{0} + (\alpha-\beta)\ket{1}$   \\
    $\mathbf{10}$  & $\mathbf{Z}$  &  $-i(\alpha-\beta)\ket{0} - (\alpha+\beta)\ket{1}$  \\
    $\mathbf{11}$  & $\mathbf{ZX}$ &  $(\alpha-\beta)\ket{0} - i(\alpha+\beta) \ket{1})$  
  \end{tabular}
\end{table}

\noindent 
As shown above, none of the standard restoration operations correctly recover the original message state $\ket{\text{Msg}} = \alpha\ket{0} + \beta\ket{1}$. This demonstrates that the restoration operations must be adapted to the measurement basis. Let us denote the appropriate restoration operators as $U_{00}, U_{01}, U_{10}, U_{11}$, corresponding to the classical outcomes of Alice’s measurement. To determine these operators, we solve the following system:

\begin{align}\label{eq:qt_set}
\begin{cases}
 U_{00}\Bigl[(\alpha + \beta)\ket{0} + i(-\alpha + \beta)\ket{1}\Bigr] 
    = \alpha\ket{0} + \beta\ket{1}, \\[6pt]
 U_{01}\Bigl[(\alpha - \beta)\ket{0} + i(\alpha + \beta)\ket{1}\Bigr] 
    = \alpha\ket{0} + \beta\ket{1}, \\[6pt]
 U_{10}\Bigl[i(\alpha - \beta\ket{0} + (\alpha + \beta)\ket{1}\Bigr] 
    = \alpha\ket{0} + \beta\ket{1}, \\[6pt]
 U_{11}\Bigl[i(\alpha + \beta)\ket{0} + (-\alpha + \beta)\ket{1}\Bigr] 
    = \alpha\ket{0} + \beta\ket{1}.
\end{cases}
\end{align}

\noindent These operators must also satisfy the constraint:

\begin{equation}
    \frac{U_{01}}{U_{00}} \cdot U_{10} = U_{11}
\label{eq:relation}
\end{equation}

\noindent Taking $U_{00}$ as an example, we solve:

\begin{align*}
    U_{00}
    \begin{pmatrix}
        \alpha + \beta\\
        i(-\alpha + \beta)
    \end{pmatrix}
    =
    \begin{pmatrix}
        \alpha\\
        \beta
    \end{pmatrix}
    \quad \text{for all } \alpha,\beta
\end{align*}

\noindent For two special cases:

\begin{align*}
    &\quad \alpha=1,\;\beta=0 \;\Rightarrow\;
        U_{00}\begin{pmatrix}1\\ -i\end{pmatrix}
        = \begin{pmatrix}1\\ 0\end{pmatrix}\\
    &\quad \alpha=0,\;\beta=1 \;\Rightarrow\;
        U_{00}\begin{pmatrix}1\\ i\end{pmatrix}
        = \begin{pmatrix}0\\ 1\end{pmatrix}
\end{align*}

\noindent Constructing a matrix $M$ with these columns:

\begin{align*}
M = \begin{pmatrix}
1 & 1 \\
-i & i
\end{pmatrix}
\quad \Rightarrow \quad
U_{00} = M^{-1} = \frac{1}{2i}
\begin{pmatrix}
i & -1 \\
i & 1
\end{pmatrix}
= \frac{1}{2}
\begin{pmatrix}
1 & i \\
1 & -i
\end{pmatrix}
\end{align*}

\noindent This matrix corresponds to the composite operation \textbf{HS}, where the global normalization factor is omitted as it does not affect the quantum state. Resolving the remaining $U$ operators yields the following:

\begin{table}[H]
  \centering
  \begin{tabular}{ccc}
    \textbf{Bits from Alice}   &  \textbf{Bob's Pre-Restoration State} &  \textbf{Restoration} \\
    \midrule
    $\mathbf{00}$  & $(\alpha+\beta)\ket{0} + i(-\alpha+\beta)\ket{1}$   &  $\mathbf{HS}$\\
    $\mathbf{01}$  & $(\alpha-\beta)\ket{0} + i(\alpha+\beta)\ket{1}$   &  $\mathbf{HSX}$\\
    $\mathbf{10}$  & $i(\alpha-\beta)\ket{0} + (\alpha+\beta)\ket{1}$  &  $\mathbf{HSY}$\\
    $\mathbf{11}$  & $i(\alpha+\beta)\ket{0} + (-\alpha+\beta)\ket{1})$  &  $\mathbf{HSYX}$
  \end{tabular}
\end{table}

\begin{figure}[htbp]
    \centering
    \includegraphics[width=\linewidth]{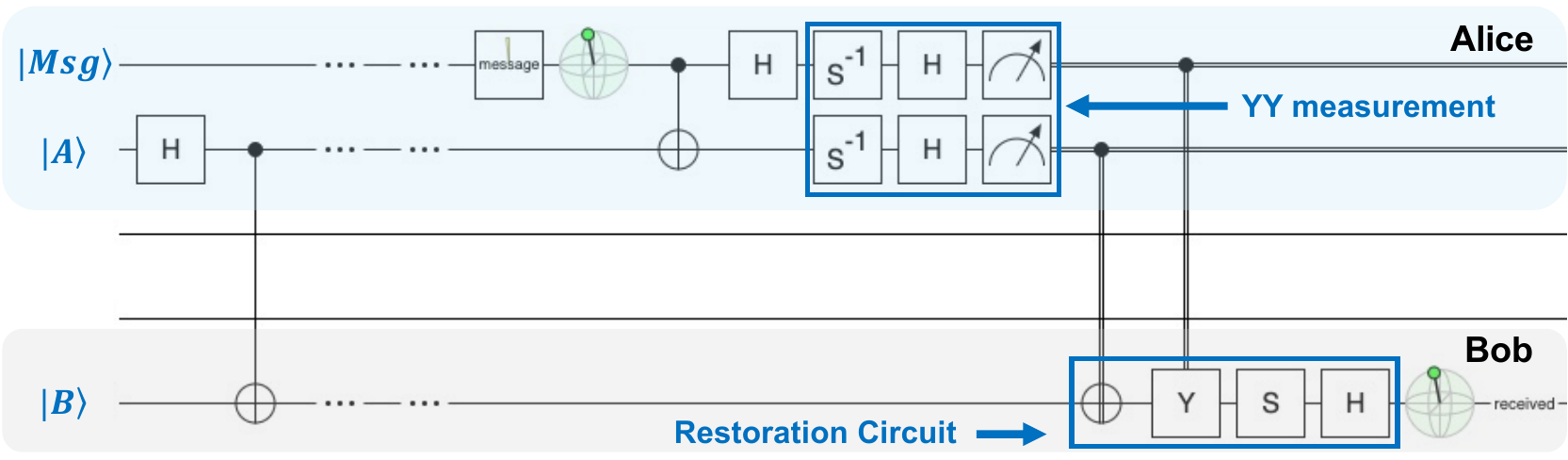}
    \caption{Quantum teleportation protocol where both $\ket{\text{Msg}}$ and $\ket{\text{A}}$ are measured in \Ybasis. Bob recovers the original message state $\alpha\ket{0}+\beta\ket{1}$ with restoration $\{\text{HS}, \text{HSX}, \text{HSY}, \text{HSYX}\}$.}
    \label{fig:QT_YY}
\end{figure}

\noindent The corrected operations ensure faithful state recovery, as illustrated in \cref{fig:QT_YY}, where Bob successfully reconstructs Alice’s original state using the updated restoration circuit.

\subsection{Hybrid Measurements}

In \cref{sec:YY}, we discussed the scenario where both the message qubit and Alice’s EPR qubit are measured in the \Ybasis. In this section, we explore \emph{hybrid measurements}, where the message qubit and Alice’s EPR qubit are measured in different bases.

\subsubsection{Teleportation with \Ybasis Measurement on $\ket{\text{Msg}}$ Only}

\begin{figure}[htbp]
    \centering
    \includegraphics[width=\linewidth]{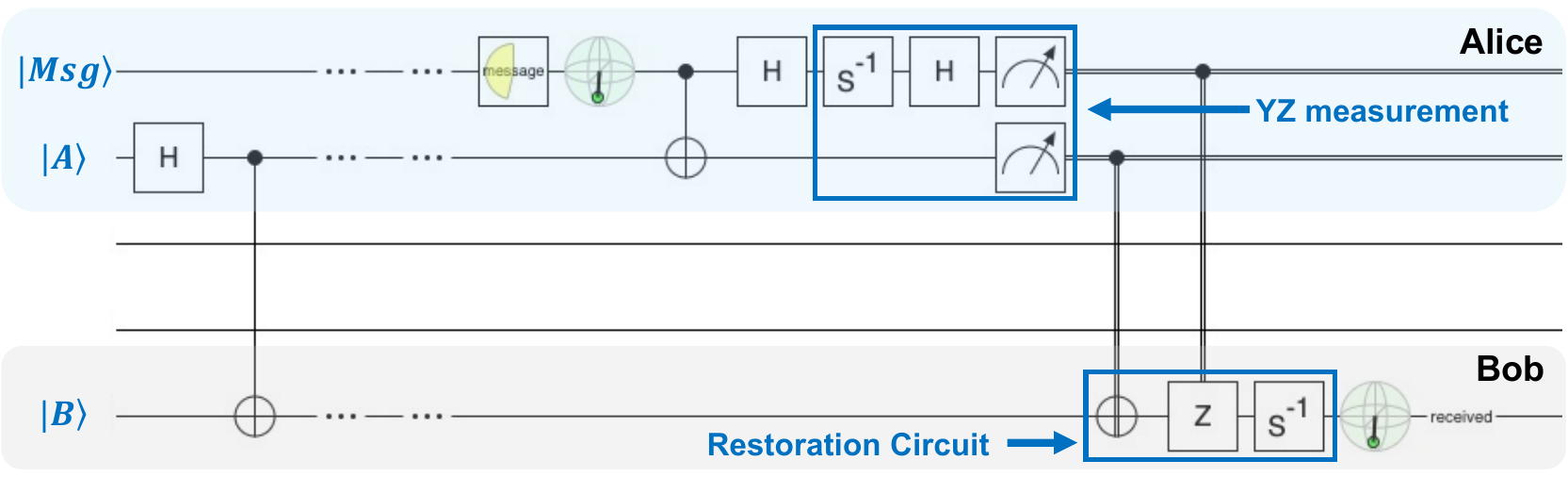}
    \caption{Quantum teleportation protocol where $\ket{\text{Msg}}$ is measured in \Ybasis and $\ket{\text{A}}$ in \Zbasis. Bob recovers the original message state with restoration: $\{\mathrm{S^\dag}, \mathrm{S^\dag X}, \mathrm{S^\dag Z}, \mathrm{S^\dag ZX}\}$.}
    \label{fig:QT_YZ}
\end{figure}

The first hybrid case that we will discuss is Y measurement only on message qubit $\ket{\text{Msg}}$, as shown in \cref{fig:QT_YZ}.

\noindent The state of the system is thus transformed to:

\begin{align*}
(H \otimes I \otimes I)& (S^\dag \otimes I \otimes I)(H\otimes I \otimes I)(\text{CNOT}\otimes I)(\ket{\text{Msg}}\otimes\ket{\text{Bell}})\\
 = \frac{\sqrt{2}(1-i)}{4} 
     & \Bigl[\quad \ket{00}(\alpha\ket{0} + i\beta\ket{1})\\
     & + \ket{01}(i\beta\ket{0} + \alpha\ket{1})\\
     & + \ket{10}(i\alpha\ket{0} + \beta\ket{1})\\
     & + \ket{11}(\beta\ket{0} + i\alpha\ket{1})\Bigr]
\end{align*}

\noindent
To correctly reconstruct the message state, we refer to \cref{eq:qt_set} and \cref{eq:relation}, then construct the following set of equations based on the updated system state. Solving this system allows us to determine the appropriate restoration operations:

\begin{align*}
\begin{cases}
 U_{00}\Bigl[\alpha \ket{0} + i\beta\ket{1}\Bigr] 
    = \alpha\ket{0} + \beta\ket{1} \\[6pt]
 U_{01}\Bigl[i\beta\ket{0} + \alpha\ket{1}\Bigr] 
    = \alpha\ket{0} + \beta\ket{1} \\[6pt]
 U_{10}\Bigl[i\alpha\ket{0} + \beta\ket{1}\Bigr] 
    = \alpha\ket{0} + \beta\ket{1} \\[6pt]
 U_{11}\Bigl[\beta\ket{0} + i\alpha\ket{1}\Bigr] 
    = \alpha\ket{0} + \beta\ket{1} \\[6pt]
 \displaystyle \frac{U_{01}}{U_{00}} \cdot U_{10} = U_{11}
\end{cases}
\end{align*}

\noindent 
Solving this equation set yields:

\begin{align*}
 U_{00}&=\begin{pmatrix} 1 & 0 \\ 0 & -i \end{pmatrix}, \quad
 U_{01} =\begin{pmatrix} 0 & 1 \\ -i & 0 \end{pmatrix} \\
 U_{10}&=\begin{pmatrix} 1 & 0 \\ 0 & i \end{pmatrix}, \quad
 U_{11} =\begin{pmatrix} 0 & 1 \\ i & 0 \end{pmatrix}
\end{align*}

\noindent
By decomposing the above $U$ matrices, we obtain the following operators that correctly transform Bob’s qubit $\ket{\text{B}}$ into the original message state $\ket{\text{Msg}}$:

\begin{table}[H]
  \centering
  \begin{tabular}{ccc}
   \textbf{Bits from Alice}   &  \textbf{Bob's Pre-Restoration State} &  \textbf{Restoration} \\
    \midrule
    $\mathbf{00}$   & $\alpha\ket{0} + i\beta\ket{1}$  & $\mathbf{S^\dag}$\\
    $\mathbf{01}$   & $i\beta\ket{0} + \alpha\ket{1}$  & $\mathbf{S^\dag X}$\\
    $\mathbf{10}$   & $i\alpha\ket{0} + \beta\ket{1}$  & $\mathbf{S^\dag Z}$\\
    $\mathbf{11}$   & $\beta\ket{0} + i\alpha\ket{1}$  & $\mathbf{S^\dag ZX}$
  \end{tabular}
\end{table}

\subsubsection{Teleportation with \Ybasis Measurement on $\ket{\text{A}}$ Only}
\begin{figure}[H]
    \centering
    \includegraphics[width=\linewidth]{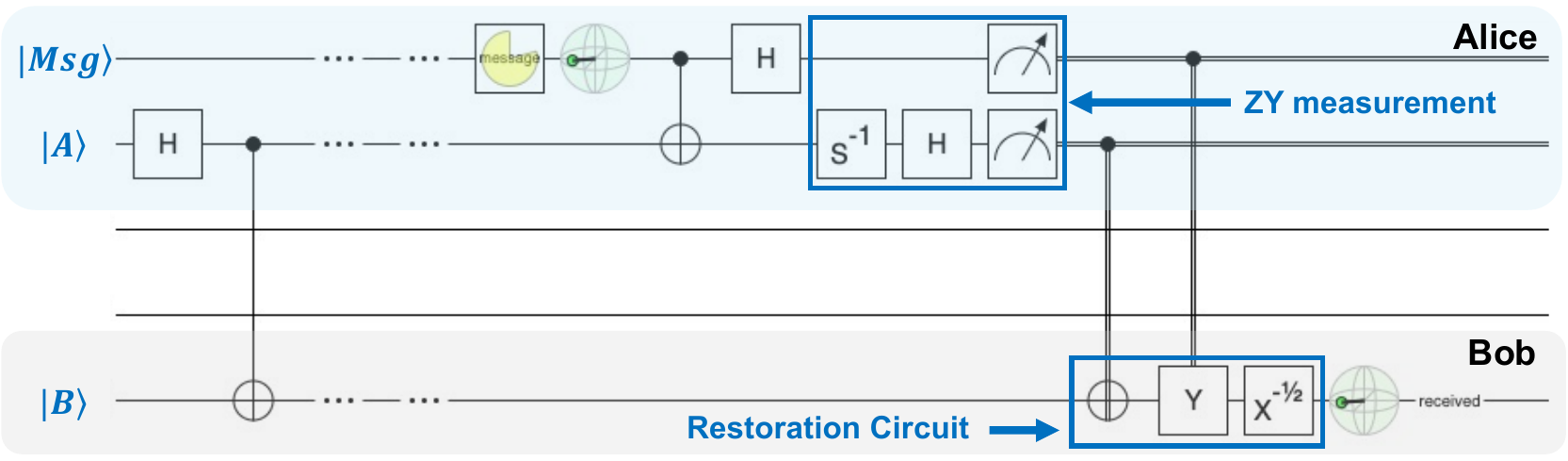}
    \caption{Quantum teleportation protocol where $\ket{\text{Msg}}$ is measured in \Zbasis and $\ket{\text{A}}$ in \Ybasis. Bob recovers the original message state with restoration: $\{\mathrm{\sqrt{X}}, \mathrm{\sqrt{X} X}, \mathrm{\sqrt{X} Y}, \mathrm{\sqrt{X} YX}\}$.}
    \label{fig:QT_ZY}
\end{figure}

In the second hybrid case, Alice measures only her EPR qubit $\ket{\text{A}}$ in the \Ybasis, while the message qubit $\ket{\text{Msg}}$ is measured in the standard \Zbasis. This circuit setup is shown in \cref{fig:QT_ZY}.

\noindent The system state becomes:
\begin{align*}
&(I \otimes H \otimes I)(I \otimes S^\dag \otimes I)(H\otimes I \otimes I)(\text{CNOT}\otimes I)(\ket{\text{Msg}}\otimes\ket{\text{Bell}})\\
 & = \frac{\sqrt{2}}{4}\Bigl[
    \quad\ket{00}[(\alpha - i\beta)\ket{0} + (-i\alpha + \beta)\ket{1}]\\
     & \qquad\quad + \ket{01}[(\alpha + i\beta)\ket{0} + (i\alpha + \beta)\ket{1}]\\
     & \qquad\quad + \ket{10}[(\alpha + i\beta)\ket{0} + (-i\alpha - \beta)\ket{1}]\\
     & \qquad\quad + \ket{11}[(\alpha - i\beta)\ket{0} + (i\alpha - \beta)\ket{1}]\Bigr]
\end{align*}

\noindent
Solving the corresponding restoration equations:

\begin{align*}
\begin{cases}
 U_{00}\Bigl[(\alpha - i\beta)\ket{0} + (-i\alpha + \beta)\ket{1}\Bigr] 
    = \alpha\ket{0} + \beta\ket{1} \\[6pt]
 U_{01}\Bigl[(\alpha + i\beta)\ket{0} + (i\alpha + \beta)\ket{1}\Bigr] 
    = \alpha\ket{0} + \beta\ket{1} \\[6pt]
 U_{10}\Bigl[(\alpha + i\beta\ket{0} + (-i\alpha - \beta)\ket{1}\Bigr] 
    = \alpha\ket{0} + \beta\ket{1} \\[6pt]
 U_{11}\Bigl[(\alpha - i\beta)\ket{0} + (i\alpha - \beta)\ket{1}\Bigr] 
    = \alpha\ket{0} + \beta\ket{1}\\[6pt]
 \displaystyle \frac{U_{01}}{U_{00}} \cdot U_{10} = U_{11}
\end{cases}
\end{align*}

\noindent
The corresponding unitary matrices are:

\begin{align*}
 U_{00}&=\begin{pmatrix} 1-i & 1+i \\ 1+i & 1-i \end{pmatrix}, \;
 U_{01} =\begin{pmatrix} 1+i & 1-i \\ 1-i & 1+i \end{pmatrix} \; \\
 U_{10}&=\begin{pmatrix} i-1 & -1-i \\ 1+i & 1-i \end{pmatrix}, \;
 U_{11} =\begin{pmatrix}-1-i & i-1 \\ 1-i & 1+i \end{pmatrix} \;
\end{align*}

\noindent
These transformations correspond to the following restoration operators by decomposing the above $U$ matrices:

\begin{table}[H]
  \centering
  \begin{tabular}{ccc}
    \textbf{Bits from Alice}   &  \textbf{Bob's Pre-Restoration State} &  \textbf{Restoration} \\
    \midrule
    $\mathbf{00}$ & $(\alpha - i\beta)\ket{0} + (-i\alpha + \beta)\ket{1}$      &  $\mathbf{\sqrt{X}^\dag}$\\
    $\mathbf{01}$ & $(\alpha + i\beta)\ket{0} + (i\alpha + \beta)\ket{1}$    &  $\mathbf{\sqrt{X}^\dag X}$\\
    $\mathbf{10}$ & $(\alpha + i\beta)\ket{0} - (i\alpha + \beta)\ket{1}$      &  $\mathbf{\sqrt{X}^\dag Y}$\\
    $\mathbf{11}$ &  $(\alpha - i\beta)\ket{0} + (i\alpha - \beta)\ket{1}$    & $\mathbf{\sqrt{X}^\dag YX}$
  \end{tabular}
\end{table}

\noindent
For the remaining hybrid measurement combinations using bases from $\{\mathrm{Y}, -\mathrm{Y}, \mathrm{Z}\}$, we summarize the corresponding pre-restoration states and restoration operations in \cref{sec:appx}.

\subsection{Teleportation with Arbitrary Measurement}
Building on the above derivations, we generalize the teleportation protocol to support arbitrary measurement bases on both the message qubit $\ket{\text{Msg}}$ and Alice’s EPR qubit $\ket{\text{A}}$. 
Let $U_{\text{msg}}$ and $U_A$ denote the unitary transformations that map the arbitrary measurement bases of the message qubit and Alice’s EPR qubit, respectively, to the computational \Zbasis:

\begin{equation*}
    U_{\text{msg}} = 
    \begin{pmatrix}
        x_{11} & x_{12} \\
        x_{21} & x_{22}
    \end{pmatrix},
    \quad
    U_{A} = 
    \begin{pmatrix}
        y_{11} & y_{12} \\
        y_{21} & y_{22}
    \end{pmatrix}
\end{equation*}

\noindent 
The system state evolves as follows:

\begin{align*}
& (U_{\text{msg}} \otimes U_A \otimes I)(H \otimes I \otimes I)(\text{CNOT} \otimes I)(\ket{\text{Msg}} \otimes \ket{\text{Bell}}) \\
&= \frac{1}{2} \Bigl[
  \ \ket{00} \bigl( \phi_{00}^{(0)} \ket{0} + \phi_{00}^{(1)} \ket{1} \bigr) \\
& \qquad + \ket{01} \bigl( \phi_{01}^{(0)} \ket{0} + \phi_{01}^{(1)} \ket{1} \bigr) \\
& \qquad + \ket{10} \bigl( \phi_{10}^{(0)} \ket{0} + \phi_{10}^{(1)} \ket{1} \bigr) \\
& \qquad + \ket{11} \bigl( \phi_{11}^{(0)} \ket{0} + \phi_{11}^{(1)} \ket{1} \bigr) \Bigr]
\end{align*}

\noindent
where the coefficients are defined as:

\begin{align*}
\phi_{00}^{(0)} &= \alpha(x_{11}y_{11} + x_{12}y_{11}) + \beta(x_{11}y_{12} - x_{12}y_{12}) \\
\phi_{00}^{(1)} &= \alpha(x_{11}y_{12} + x_{12}y_{12}) + \beta(x_{11}y_{11} - x_{12}y_{11}) \\
\phi_{01}^{(0)} &= \alpha(x_{11}y_{21} + x_{12}y_{21}) + \beta(x_{11}y_{22} - x_{12}y_{22}) \\
\phi_{01}^{(1)} &= \alpha(x_{11}y_{22} + x_{12}y_{22}) + \beta(x_{11}y_{21} - x_{12}y_{21}) \\
\phi_{10}^{(0)} &= \alpha(x_{21}y_{11} + x_{22}y_{11}) + \beta(x_{21}y_{12} - x_{22}y_{12}) \\
\phi_{10}^{(1)} &= \alpha(x_{21}y_{12} + x_{22}y_{12}) + \beta(x_{21}y_{11} - x_{22}y_{11}) \\
\phi_{11}^{(0)} &= \alpha(x_{21}y_{21} + x_{22}y_{21}) + \beta(x_{21}y_{22} - x_{22}y_{22}) \\
\phi_{11}^{(1)} &= \alpha(x_{21}y_{22} + x_{22}y_{22}) + \beta(x_{21}y_{21} - x_{22}y_{21})
\end{align*}

\noindent The corresponding restoration operations must satisfy:

\begin{align*}
\begin{cases}
U_{00} \left( \phi_{00}^{(0)} \ket{0} + \phi_{00}^{(1)} \ket{1} \right) = \alpha\ket{0} + \beta\ket{1} \\[6pt]
U_{01} \left( \phi_{01}^{(0)} \ket{0} + \phi_{01}^{(1)} \ket{1} \right) = \alpha\ket{0} + \beta\ket{1} \\[6pt]
U_{10} \left( \phi_{10}^{(0)} \ket{0} + \phi_{10}^{(1)} \ket{1} \right) = \alpha\ket{0} + \beta\ket{1} \\[6pt]
U_{11} \left( \phi_{11}^{(0)} \ket{0} + \phi_{11}^{(1)} \ket{1} \right) = \alpha\ket{0} + \beta\ket{1} \\[6pt]
\displaystyle \frac{U_{01}}{U_{00}} \cdot U_{10} = U_{11}
\end{cases}
\end{align*}
\noindent
This formulation ensures that teleportation remains correct under arbitrary local basis choices.

\section{Discussion}\label{sec:discussion}
This section explores the security implications of our proposed multi-axis quantum teleportation protocol. While canonical teleportation prevents external eavesdropping by not transmitting the quantum state directly, it assumes honest participants and secure entanglement distribution. This assumption exposes the protocol to insider threats and adversary-in-the-middle (AitM) attacks \cite{aitm}, as it lacks mechanisms to authenticate participants or verify entanglement ownership. For instance, an adversary who intercepts Bob’s entangled qubit and later obtains Alice’s classical measurement outcome can reconstruct the teleported state, thereby compromising the protocol. The standard formulation offers no intrinsic defenses against such misuse \cite{qt, qt_progress}.

Our proposed modification addresses this vulnerability by dynamically varying the sender’s measurement basis rather than restricting it to the fixed $Z$-basis. By selecting the basis from a secret or randomly chosen parameter, the protocol gains an encryption-like property: even if an adversary obtains both the entangled qubit and classical bits, they cannot reconstruct the state without knowing the correct basis. Legitimate participants can coordinate basis choices through pre-shared secrets or encrypted classical messages to ensure proper decoding. Any unauthorized interception or incorrect basis use results in a corrupted state, effectively nullifying the attack. This approach aligns with findings that measurement dependence can enhance security in quantum networks \cite{Kundu2024MeasurementDependence}.

While reminiscent of controlled teleportation schemes that rely on third-party authorization \cite{Gao2004ControlledQT, Zhou2020ExperimentalCQT}, our approach achieves similar security benefits without additional infrastructure. Making the measurement basis adaptive and private enables the protocol to be more resilient to insider threats and distribution attacks. Although a full cryptographic analysis is beyond this work's scope, the proposed method significantly limits an adversary’s ability to recover the teleported state.

\section{Conclusion}
In this paper, we present a generalized quantum teleportation framework that systematically incorporates diverse measurement bases. We provide self-contained proofs to establish the restorations under different measurement bases. This work lays the groundwork for flexible and secure teleportation schemes, advancing quantum information communication.

\section{APPENDIX} \label{sec:appx}

\subsection{\mYbasis Measurement on Both $\ket{\text{Msg}}$ and $\ket{\text{A}}$}
\noindent The state of the system is transformed to:
\begin{align*}
&(H \otimes H \otimes I)(S \otimes S \otimes I)(H\otimes I \otimes I)(\text{CNOT}\otimes I)(\ket{\text{Msg}}\otimes\ket{\text{Bell}})\\
 &    = \frac{1-i}{4}\Bigl[
                     \ket{00}[i(\alpha+\beta)\ket{0} + (-\alpha+\beta)\ket{1}]\\
     & \qquad\quad + \ket{01}[i(\alpha-\beta)\ket{0} + (\alpha+\beta)\ket{1}]\\
     & \qquad\quad + \ket{10}[(\alpha-\beta)\ket{0} + i(\alpha+\beta)\ket{1}]\\
     & \qquad\quad - \ket{11}[(\alpha+\beta)\ket{0} + i(-\alpha+\beta)\ket{1}]\Bigr]
\end{align*}

\begin{figure}[H]
    \centering
    \includegraphics[width=\linewidth]{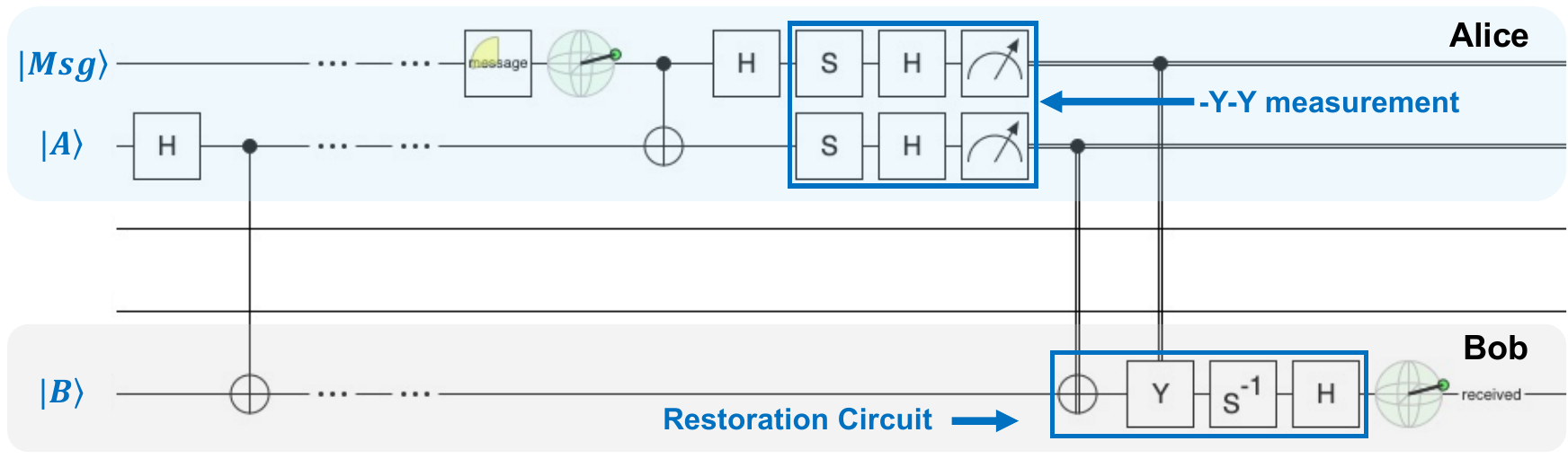}
    \caption{Quantum teleportation protocol where $\ket{\text{Msg}}$ and $\ket{\text{A}}$ are both measured in \mYbasis. Bob recovers the original message state with restoration: $\{\mathrm{HS^\dag}, \mathrm{HS^\dag X}, \mathrm{HS^\dag Y}, \mathrm{HS^\dag YX}\}$.}
    \label{fig:QT_-Y-Y}
\end{figure}

\noindent The updated restorations are illustrated in \cref{fig:QT_-Y-Y}. With the below operators, Bob can appropriately transform his EPR qubit to the message state:
\begin{table}[H]
  \centering
  \begin{tabular}{ccc}
    \textbf{Bits from Alice}   &  \textbf{Bob's Pre-Restoration State} &  \textbf{Restoration} \\
    \midrule
    $\mathbf{00}$  & $i(\alpha+\beta)\ket{0} + (-\alpha+\beta)\ket{1}$  &  $\mathbf{HS^\dag}$\\
    $\mathbf{01}$  & $i(\alpha-\beta)\ket{0} + (\alpha+\beta)\ket{1}$   &  $\mathbf{HS^\dag X}$\\
    $\mathbf{10}$  & $(\alpha-\beta)\ket{0} + i(\alpha+\beta)\ket{1}$   &  $\mathbf{HS^\dag Y}$\\
    $\mathbf{11}$  & $(\alpha+\beta)\ket{0} + i(-\alpha+\beta)\ket{1})$ &  $\mathbf{HS^\dag YX}$
  \end{tabular}
\end{table}


\subsection{\mYbasis Measurement on $\ket{\text{Msg}}$ Only}
\noindent The state of the system is transformed to:
\begin{align*}
&(H \otimes I \otimes I)(S \otimes I \otimes I)(H\otimes I \otimes I)(\text{CNOT}\otimes I)(\ket{\text{Msg}}\otimes\ket{\text{Bell}})\\
 &    = \frac{\sqrt{2}(1-i)}{4}\Bigl[
                        \ \,\ket{00}(\:i\alpha\ket{0} + \beta\ket{1}\:)\\
     & \qquad\qquad\qquad + \ket{01}(\:\beta\ket{0} + i\alpha\ket{1}\:)\\
     & \qquad\qquad\qquad + \ket{10}(\:\alpha\ket{0} + i\beta\ket{1}\:)\\
     & \qquad\qquad\qquad + \ket{11}(\:i\beta\ket{0} + \alpha\ket{1}\:)\Bigr]
\end{align*}

\begin{figure}[H]
    \centering
    \includegraphics[width=\linewidth]{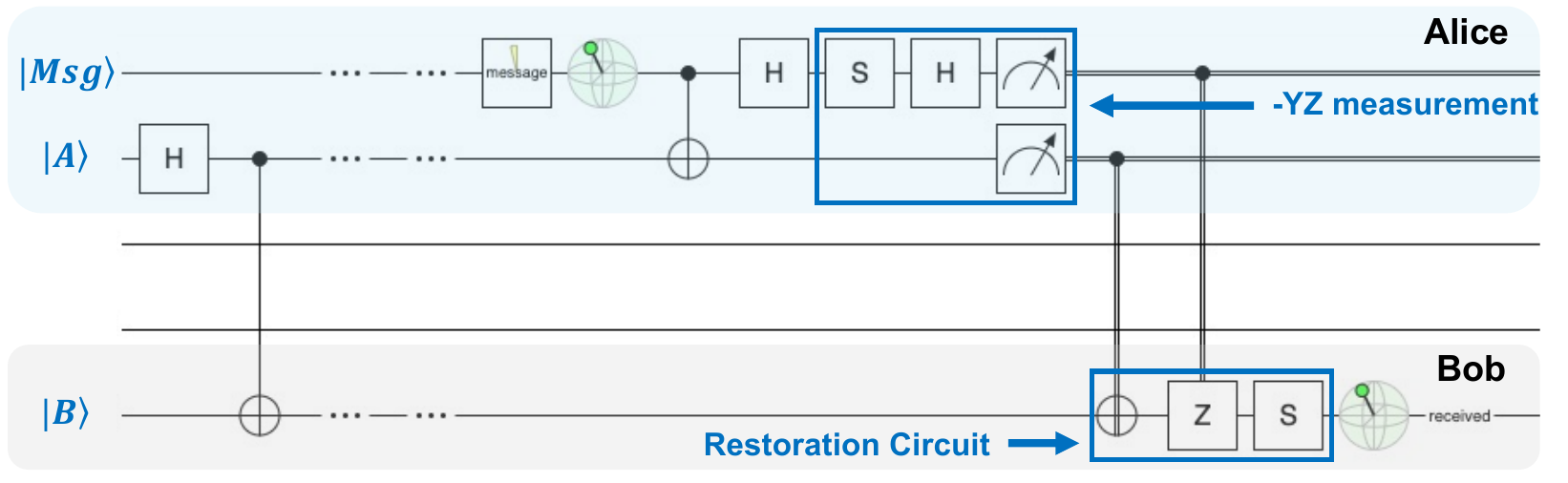}
    \caption{Quantum teleportation protocol where $\ket{\text{Msg}}$ is measured in \mYbasis and $\ket{\text{A}}$ in \Zbasis. Bob recovers the original message state with restoration: $\{\mathrm{S}, \mathrm{SX}, \mathrm{SZ}, \mathrm{SZX}\}$.}
    \label{fig:QT_-YZ}
\end{figure}

\noindent The updated restorations are illustrated in \cref{fig:QT_-YZ}. With the below operators, Bob can appropriately recover the message:
\begin{table}[H]
  \centering
  \begin{tabular}{ccc}
    \textbf{Bits from Alice}   &  \textbf{Bob's Pre-Restoration State} &  \textbf{Restoration}\\
    \midrule
    $\mathbf{00}$  & $i\alpha\ket{0} + \beta\ket{1}$ &  $\mathbf{S}$\\
    $\mathbf{01}$  & $\beta\ket{0} + i\alpha\ket{1}$ &  $\mathbf{SX}$\\
    $\mathbf{10}$  & $\alpha\ket{0} + i\beta\ket{1}$ &  $\mathbf{SZ}$\\
    $\mathbf{11}$  & $i\beta\ket{0} + \alpha\ket{1}$ &  $\mathbf{SZX}$
  \end{tabular}
\end{table}

\subsection{\mYbasis Measurement on $\ket{\text{A}}$ Only}
\noindent The state of the system is transformed to:
\begin{align*}
&(I \otimes H \otimes I)(I \otimes S \otimes I)(H\otimes I \otimes I)(\text{CNOT}\otimes I)(\ket{\text{Msg}}\otimes\ket{\text{Bell}})\\
 &    = \frac{\sqrt{2}}{4}\Bigl[
                \ \, \ket{00}[\:(\alpha+i\beta)\ket{0} + (i\alpha+\beta)\ket{1} \:]\\
     & \qquad\quad + \ket{01}[\:(\alpha-i\beta)\ket{0} + (-i\alpha+\beta)\ket{1}\:]\\
     & \qquad\quad + \ket{10}[\:(\alpha-i\beta)\ket{0} + (i\alpha-\beta)\ket{1} \:]\\
     & \qquad\quad - \ket{11}[\:(\alpha+i\beta)\ket{0} + (-i\alpha-\beta)\ket{1}\:]\Bigr]
\end{align*}

\begin{figure}[H]
    \centering
    \includegraphics[width=\linewidth]{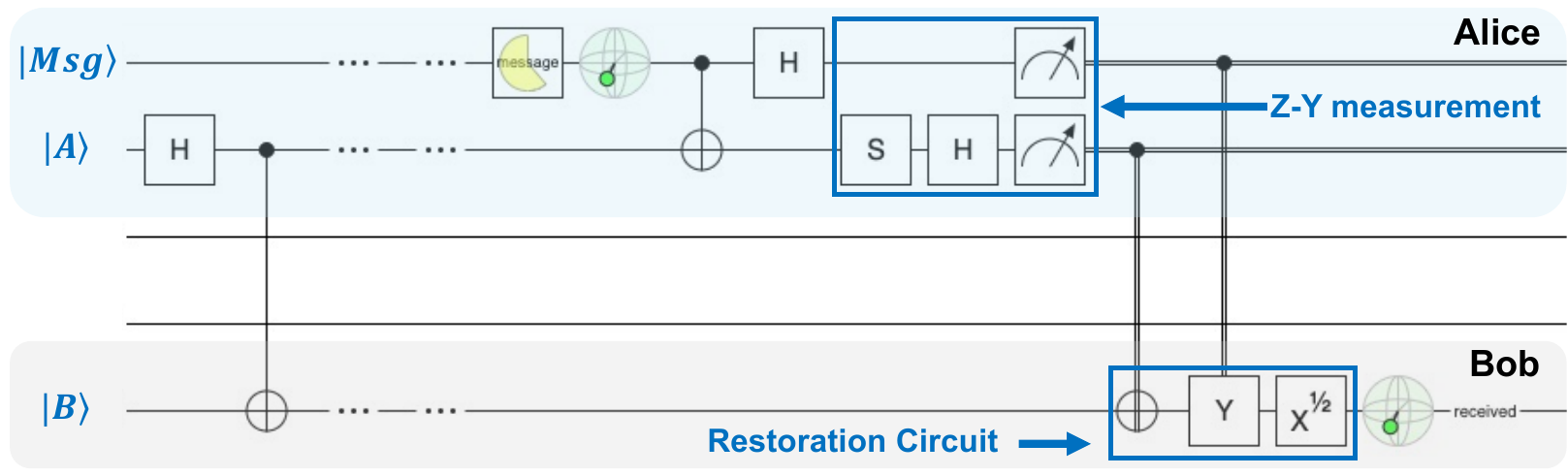}
    \caption{Quantum teleportation protocol where $\ket{\text{Msg}}$ is measured in \Zbasis and $\ket{\text{A}}$ in \mYbasis. Bob recovers the original message state with restoration: $\{\mathrm{\sqrt{X}}, \mathrm{\sqrt{X} X}, \mathrm{\sqrt{X} Z}, \mathrm{\sqrt{X} ZX}\}$.}
    \label{fig:QT_Z-Y}
\end{figure}

\noindent The updated restorations are illustrated in \cref{fig:QT_Z-Y}. With the below operators, Bob can appropriately recover the message:
\begin{table}[H]
  \centering
  \begin{tabular}{ccc}
    \textbf{Bits from Alice}   &  \textbf{Bob's Pre-Restoration State} &  \textbf{Restoration} \\
    \midrule
    $\mathbf{00}$  & $(\alpha+i\beta)\ket{0} + (i\alpha+\beta)\ket{1}$   &  $\mathbf{\sqrt{X}}$\\
    $\mathbf{01}$  & $(\alpha-i\beta)\ket{0} + (-i\alpha+\beta)\ket{1}$ &  $\mathbf{\sqrt{X}X}$\\
    $\mathbf{10}$  & $(\alpha-i\beta)\ket{0} + (i\alpha-\beta)\ket{1}$  &  $\mathbf{\sqrt{X}Y}$\\
    $\mathbf{11}$  & $(\alpha+i\beta)\ket{0} + (-i\alpha-\beta)\ket{1}$ &  $\mathbf{\sqrt{X}YX}$
  \end{tabular}
\end{table}

\subsection{\mYbasis Measurement on $\ket{\text{Msg}}$ and \Ybasis on $\ket{\text{A}}$}
\noindent The state of the system is transformed to:
\begin{align*}
&(H \otimes H \otimes I)(S \otimes S^\dag \otimes I)(H\otimes I \otimes I)(\text{CNOT}\otimes I)(\ket{\text{Msg}}\otimes\ket{\text{Bell}})\\
 &   = \frac{1-i}{4}\Bigl[
                     \ket{00}[\:i(\alpha-\beta)\ket{0} + (\alpha+\beta)\ket{1} \:]\\
     & \qquad\quad + \ket{01}[\:i(\alpha+\beta)\ket{0} + (-\alpha+\beta)\ket{1}\:]\\
     & \qquad\quad + \ket{10}[\:(\alpha+\beta)\ket{0} + i(-\alpha+\beta)\ket{1}\:]\\
     & \qquad\quad - \ket{11}[\:(\alpha-\beta)\ket{0} + i(\alpha+\beta)\ket{1} \:]\Bigr]
\end{align*}

\begin{figure}[H]
    \centering
    \includegraphics[width=\linewidth]{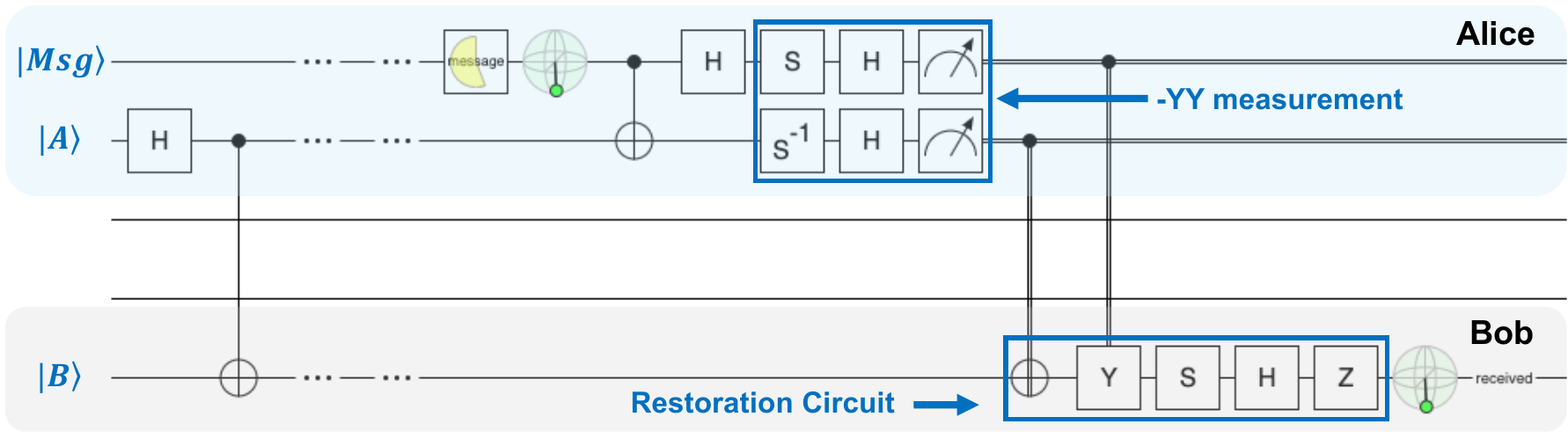}
    \caption{Quantum teleportation protocol where $\ket{\text{Msg}}$ is measured in \mYbasis and $\ket{\text{A}}$ in \Ybasis. Bob recovers the original message state with restoration: $\{\mathrm{ZHS}, \mathrm{ZHS X}, \mathrm{ZHS Y}, \mathrm{ZHS YX}\}$.}
    \label{fig:QT_-YY}
\end{figure}

\noindent The updated restorations are illustrated in \cref{fig:QT_-YY}. With the below operators, Bob can appropriately recover the message:
\begin{table}[H]
  \centering
  \begin{tabular}{ccc}
   \textbf{Bits from Alice}   &  \textbf{Bob's Pre-Restoration State} &  \textbf{Restoration} \\
    \midrule
    $\mathbf{00}$  & $i(\alpha-\beta)\ket{0} + (\alpha+\beta)\ket{1}$  &  $\mathbf{ZHS}$\\
    $\mathbf{01}$  & $i(\alpha+\beta)\ket{0} + (-\alpha+\beta)\ket{1}$ &  $\mathbf{ZHS X}$\\
    $\mathbf{10}$  & $(\alpha+\beta)\ket{0} + i(-\alpha+\beta)\ket{1}$ &  $\mathbf{ZHS Y}$\\
    $\mathbf{11}$  & $(\alpha-\beta)\ket{0} + i(\alpha+\beta)\ket{1})$ &  $\mathbf{ZHS YX}$
  \end{tabular}
\end{table}

\subsection{\Ybasis Measurement on $\ket{\text{Msg}}$ and \mYbasis on $\ket{\text{A}}$}
\noindent The state of the system is transformed to:
\begin{align*}
&(H \otimes H \otimes I)(S^\dag \otimes S \otimes I)(H\otimes I \otimes I)(\text{CNOT}\otimes I)(\ket{\text{Msg}}\otimes\ket{\text{Bell}})\\
 &   = \frac{1-i}{4}\Bigl[
                     \ket{00}[\;(\alpha-\beta)\ket{0} + i(\alpha+\beta)\ket{1}\;]\\
     & \qquad\quad + \ket{01}[\;(\alpha+\beta)\ket{0} + i(-\alpha+\beta)\ket{1}\;]\\
     & \qquad\quad + \ket{10}[\;i(\alpha+\beta)\ket{0} + (-\alpha+\beta)\ket{1}\;]\\
     & \qquad\quad - \ket{11}[\;i(\alpha-\beta)\ket{0} + (\alpha+\beta)\ket{1}\;]\Bigr]
\end{align*}

\begin{figure}[H]
    \centering
    \includegraphics[width=\linewidth]{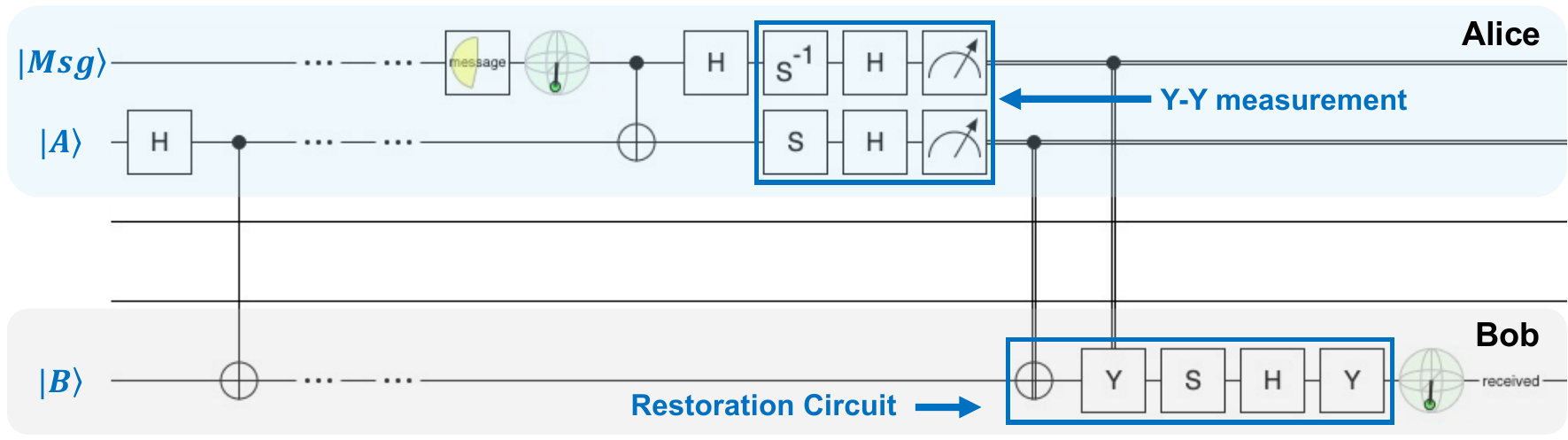}
    \caption{Quantum teleportation protocol where $\ket{\text{Msg}}$ is measured in \Ybasis and $\ket{\text{A}}$ in \mYbasis. Bob recovers the original message state with restoration: $\{\mathrm{YHS}, \mathrm{YHS X}, \mathrm{YHS Y}, \mathrm{YHS YX}\}$.}
    \label{fig:QT_Y-Y}
\end{figure}

\noindent The updated restorations are illustrated in \cref{fig:QT_Y-Y}. With the below operators, Bob can appropriately recover the message:
\begin{table}[H]
  \centering
  \begin{tabular}{ccc}
   \textbf{Bits from Alice}   &  \textbf{Bob's Pre-Restoration State} &  \textbf{Restoration} \\
    \midrule
    $\mathbf{00}$  & $(\alpha-\beta)\ket{0} + i(\alpha+\beta)\ket{1}$  &  $\mathbf{YHS}$\\
    $\mathbf{01}$  & $(\alpha+\beta)\ket{0} + i(-\alpha+\beta)\ket{1}$ &  $\mathbf{YHS X}$\\
    $\mathbf{10}$  & $i(\alpha+\beta)\ket{0} + (-\alpha+\beta)\ket{1}$ &  $\mathbf{YHS Y}$\\
    $\mathbf{11}$  & $i(\alpha-\beta)\ket{0} + (\alpha+\beta)\ket{1})$ &  $\mathbf{YHS YX}$
  \end{tabular}
\end{table}

\subsection{\Xbasis Measurement on Both $\ket{\text{Msg}}$ and $\ket{\text{A}}$}
\noindent The state of the system is transformed to:
\begin{align*}
&(H \otimes H \otimes I)(H\otimes I \otimes I)(CNOT\otimes I)(\ket{Secret}\otimes\ket{Bell})\\
 &    = \frac{1}{2}( \ket{00}(\alpha\ket{0}  + \alpha\ket{1})\\
     & \quad + \ket{01}(\alpha\ket{0} -\alpha\ket{1})\\
     & \quad + \ket{10}(\beta\ket{0} +\beta\ket{1})\\
     & \quad - \ket{11}(-\beta\ket{0} +\beta\ket{1}))
\end{align*}

\noindent
In this case, the resulting states do not admit a transformation that maps them to the original message state. Thus, the teleportation fails under this measurement configuration.


\bibliographystyle{plain}
\bibliography{reference}

@misc{Qiskit,
   title = {Qiskit},
   url= {https://www.ibm.com/quantum/qiskit},
   year = {2025},
   Author = {IBM}
}

@article{qt,
  title={Teleporting an unknown quantum state via dual classical and Einstein-Podolsky-Rosen channels},
  author={Bennett, Charles H and Brassard, Gilles and Cr{\'e}peau, Claude and Jozsa, Richard and Peres, Asher and Wootters, William K},
  journal={Physical review letters},
  volume={70},
  number={13},
  pages={1895},
  year={1993},
  publisher={APS}
}

@article{qt_progress,
  title={Progress in quantum teleportation},
  author={Hu, Xiao-Min and Guo, Yu and Liu, Bi-Heng and Li, Chuan-Feng and Guo, Guang-Can},
  journal={Nature Reviews Physics},
  volume={5},
  number={6},
  pages={339--353},
  year={2023},
  publisher={Nature Publishing Group UK London}
}

@misc{quirk,
  title={Quirk: A drag-and-drop quantum circuit simulator},
  author={Craig Gidney},
  year={2016},
  url={https://algassert.com/quirk}
}

@article{non_clone,
  title={A single quantum cannot be cloned},
  author={Wootters, William K and Zurek, Wojciech H},
  journal={Nature},
  volume={299},
  number={5886},
  pages={802--803},
  year={1982},
  publisher={Nature Publishing Group UK London}
}

@book{q_info,
  title={Quantum computation and quantum information},
  author={Nielsen, Michael A and Chuang, Isaac L},
  year={2010},
  publisher={Cambridge university press}
}

@article{quantum_internet,
  title={The quantum internet},
  author={Kimble, H Jeff},
  journal={Nature},
  volume={453},
  number={7198},
  pages={1023--1030},
  year={2008},
  publisher={Nature Publishing Group}
}

@misc{QFI_qt_2,
      title={Teleportation two-qubit state by using two different protocols}, 
      author={K. El Anouz and A. El Allati and N. Metwally},
      year={2024},
      eprint={2404.00512},
      archivePrefix={arXiv},
      primaryClass={quant-ph},
      url={https://arxiv.org/abs/2404.00512}, 
}

@inproceedings{qce_qt,
  title={Role of Error Syndromes in Teleportation Scheduling},
  author={Chandra, Aparimit and Rozpedek, Filip and Towsley, Don},
  booktitle={2024 IEEE International Conference on Quantum Computing and Engineering (QCE)},
  volume={1},
  pages={1931--1937},
  year={2024},
  organization={IEEE}
}

@misc{qt_1400,
      title={QEYSSat 2.0 -- White Paper on Satellite-based Quantum Communication Missions in Canada}, 
      author={Thomas Jennewein and Christoph Simon and Andre Fougeres and Francois Babin and Faezeh Kimiaee Asadi and others},
      year={2024},
      eprint={2306.02481},
      archivePrefix={arXiv},
      primaryClass={quant-ph},
      url={https://arxiv.org/abs/2306.02481}, 
}

@article{qt_dqc,
  title={Distributed quantum computing across an optical network link},
  author={Main, D and Drmota, P and Nadlinger, DP and Ainley, EM and Agrawal, A and Nichol, BC and Srinivas, R and Araneda, G and Lucas, DM},
  journal={Nature},
  pages={1--6},
  year={2025},
  publisher={Nature Publishing Group UK London}
}

@online{multi_axis_0,
  title     = {Pauli measurements - Azure Quantum | Microsoft Learn},
  author    = {{Microsoft}},
  year      = {2024},
  url       = {https://learn.microsoft.com/en-us/azure/quantum/concepts-pauli-measurements},
  note      = {Accessed: 2025-03-28}
}

@article{hybrid_q_photon,
  title = {Quantum teleportation of hybrid qubits and single-photon qubits using Gaussian resources},
  author = {Bose, Soumyakanti and Jeong, Hyunseok},
  journal = {Phys. Rev. A},
  volume = {105},
  issue = {3},
  pages = {032434},
  numpages = {11},
  year = {2022},
  month = {Mar},
  publisher = {American Physical Society},
  doi = {10.1103/PhysRevA.105.032434},
  url = {https://link.aps.org/doi/10.1103/PhysRevA.105.032434}
}

@article{qt_photon,
  title={Qubit teleportation between a memory-compatible photonic time-bin qubit and a solid-state quantum network node},
  author={Iuliano, Mariagrazia and Slater, Marie-Christine and Stolk, Arian J and others},
  journal={npj Quantum Information},
  volume={10},
  number={1},
  pages={107},
  year={2024},
  publisher={Nature Publishing Group UK London}
}

@article{exp_qt,
  title={Experimental quantum teleportation},
  author={Bouwmeester, Dik and Pan, Jian-Wei and Mattle, Klaus and Eibl, Manfred and Weinfurter, Harald and Zeilinger, Anton},
  journal={Nature},
  volume={390},
  number={6660},
  pages={575--579},
  year={1997},
  publisher={Nature Publishing Group UK London}
}

@misc{aitm,
  author = {{MITRE}},
  title = {Adversary-in-the-Middle, Technique {T1557} - {Enterprise}},
  year = {2024},
  url = {https://attack.mitre.org/techniques/T1557/},
}

@article{wehner_quantum_2018,
	title = {Quantum internet: {A} vision for the road ahead},
	volume = {362},
	shorttitle = {Quantum internet},
	url = {https://www.science.org/doi/10.1126/science.aam9288},
	doi = {10.1126/science.aam9288},
	number = {6412},
	urldate = {2024-04-08},
	journal = {Science},
	author = {Wehner, Stephanie and Elkouss, David and Hanson, Ronald},
	month = oct,
	year = {2018},
	note = {Publisher: American Association for the Advancement of Science},
	pages = {eaam9288},
}

@article{hermans_qubit_2022,
	title = {Qubit teleportation between non-neighbouring nodes in a quantum network},
	volume = {605},
	copyright = {2022 The Author(s)},
	issn = {1476-4687},
	url = {https://www.nature.com/articles/s41586-022-04697-y},
	doi = {10.1038/s41586-022-04697-y},
	language = {en},
	number = {7911},
	urldate = {2024-04-08},
	journal = {Nature},
	author = {Hermans, S. L. N. and Pompili, M. and Beukers, H. K. C. and Baier, S. and Borregaard, J. and Hanson, R.},
	month = may,
	year = {2022},
	note = {Publisher: Nature Publishing Group},
	pages = {663--668},
}

@article{bell_measurement,
	title = {Bell measurements for teleportation},
	volume = {59},
	url = {https://link.aps.org/doi/10.1103/PhysRevA.59.3295},
	doi = {10.1103/PhysRevA.59.3295},
	number = {5},
	urldate = {2024-04-08},
	journal = {Physical Review A},
	author = {Lütkenhaus, N. and Calsamiglia, J. and Suominen, K.-A.},
	month = may,
	year = {1999},
	note = {Publisher: American Physical Society},
	pages = {3295--3300},
}

@article{Kundu2024MeasurementDependence,
  title={Measurement dependence can enhance security in a quantum network},
  author={Amit Kundu and Debasis Sarkar},
  journal={arXiv preprint arXiv:2405.12379},
  year={2024},
  url={https://arxiv.org/abs/2405.12379}
}

@article{Zhou2020ExperimentalCQT,
  title={Experimental realization of controlled quantum teleportation},
  author={Nan Zhou and Yu-Bo Sheng},
  journal={Scientific Reports},
  volume={10},
  pages={12688},
  year={2020},
  publisher={Nature Publishing Group},
  url={https://www.nature.com/articles/s41598-020-70446-8}
}

@article{Gao2004ControlledQT,
  title={Controlled quantum teleportation and secure direct communication},
  author={Ting Gao and Fengli Yan and Zhi-xi Wang},
  journal={arXiv preprint quant-ph/0403155},
  year={2004},
  url={https://arxiv.org/abs/quant-ph/0403155}
}

@article{multi_measurement,
  title = {Asymptotic Teleportation Scheme as a Universal Programmable Quantum Processor},
  author = {Ishizaka, Satoshi and Hiroshima, Tohya},
  journal = {Phys. Rev. Lett.},
  volume = {101},
  issue = {24},
  pages = {240501},
  numpages = {4},
  year = {2008},
  month = {Dec},
  publisher = {American Physical Society},
  doi = {10.1103/PhysRevLett.101.240501},
  url = {https://link.aps.org/doi/10.1103/PhysRevLett.101.240501}
}

@article{multi_measurement_1,
  title = {General properties of nonsignaling theories},
  author = {Masanes, Ll. and Acin, A. and Gisin, N.},
  journal = {Phys. Rev. A},
  volume = {73},
  issue = {1},
  pages = {012112},
  numpages = {9},
  year = {2006},
  month = {Jan},
  publisher = {American Physical Society},
  doi = {10.1103/PhysRevA.73.012112},
  url = {https://link.aps.org/doi/10.1103/PhysRevA.73.012112}
}

@article{zhao2004experimental,
  title={Experimental demonstration of five-photon entanglement and open-destination teleportation},
  author={Zhao, Zhi and Chen, Yu-Ao and Zhang, An-Ning and Yang, Tao and Briegel, Hans J and Pan, Jian-Wei},
  journal={Nature},
  volume={430},
  number={6995},
  pages={54--58},
  year={2004},
  publisher={Nature Publishing Group UK London}
}

@article{multi_measurement_2,
  title = {Manipulating Biphotonic Qutrits},
  author = {Lanyon, B. P. and Weinhold, T. J. and Langford, N. K. and O'Brien, J. L. and Resch, K. J. and Gilchrist, A. and White, A. G.},
  journal = {Phys. Rev. Lett.},
  volume = {100},
  issue = {6},
  pages = {060504},
  numpages = {4},
  year = {2008},
  month = {Feb},
  publisher = {American Physical Society},
  doi = {10.1103/PhysRevLett.100.060504},
  url = {https://link.aps.org/doi/10.1103/PhysRevLett.100.060504}
}

@article{mor1999teleportation,
  title={Teleportation via generalized measurements, and conclusive teleportation},
  author={Mor, Tal and Horodecki, Pawel},
  journal={arXiv preprint quant-ph/9906039},
  year={1999}
}

@article{gordon2006generalized,
  title={Generalized teleportation protocol},
  author={Gordon, Goren and Rigolin, Gustavo},
  journal={Physical Review A—Atomic, Molecular, and Optical Physics},
  volume={73},
  number={4},
  pages={042309},
  year={2006},
  publisher={APS}
}

\vfill

\end{document}